\documentclass[%
reprint,
superscriptaddress,
showpacs,
 amsmath,amssymb,
 aps,
 prl,
floatfix,
]{revtex4-1}

\usepackage{graphicx}
\usepackage{dcolumn}
\usepackage{bm}

\usepackage{color}

\begin{document}


\title{Measurements of Double-Polarized Compton Scattering Asymmetries and Extraction of the Proton Spin Polarizabilities}

\author{P. P. Martel}
\email{martel@kph.uni-mainz.de}
\affiliation{Department of Physics, University of Massachusetts Amherst, Amherst, Massachusetts 01003, USA}
\affiliation{Institut f\"{u}r Kernphysik, Universit\"{a}t Mainz, D-55099 Mainz, Germany}
\affiliation{Department of Physics, Mount Allison University, Sackville, New Brunswick E4L 1E6, Canada}
\author{R. Miskimen}
\email{miskimen@physics.umass.edu}
\affiliation{Department of Physics, University of Massachusetts Amherst, Amherst, Massachusetts 01003, USA}
\author{P. Aguar-Bartolome}
\affiliation{Institut f\"{u}r Kernphysik, Universit\"{a}t Mainz, D-55099 Mainz, Germany}
\author{J. Ahrens}
\affiliation{Institut f\"{u}r Kernphysik, Universit\"{a}t Mainz, D-55099 Mainz, Germany}
\author{C. S. Akondi}
\affiliation{Department of Physics, Kent State University, Kent, Ohio 44242, USA}
\author{J. R. M. Annand}
\affiliation{SUPA School of Physics and Astronomy, University of Glasgow, Glasgow G12 8QQ, United Kingdom}
\author{H. J. Arends}
\affiliation{Institut f\"{u}r Kernphysik, Universit\"{a}t Mainz, D-55099 Mainz, Germany}
\author{W. Barnes}
\affiliation{Department of Physics, University of Massachusetts Amherst, Amherst, Massachusetts 01003, USA}
\author{R. Beck}
\affiliation{Helmholtz-Institut f\"{u}r Strahlen- und Kernphysik, Universit\"{a}t Bonn, D-53115 Bonn, Germany}
\author{A. Bernstein}
\affiliation{Laboratory for Nuclear Science, Massachusetts Institute of Technology, Cambridge, Massachusetts 02139, USA}
\author{N. Borisov}
\affiliation{Joint Institute for Nuclear Research (JINR), 141980 Dubna, Russia}
\author{A. Braghieri}
\affiliation{INFN Sezione di Pavia, I-27100 Pavia, Italy}
\author{W. J. Briscoe}
\affiliation{Department of Physics, The George Washington University, Washington, D.C. 20052, USA}
\author{S. Cherepnya}
\affiliation{Lebedev Physical Institute, 119991 Moscow, Russia}
\author{C. Collicott}
\affiliation{Department of Physics and Atmospheric Science, Dalhousie University, Halifax, Nova Scotia B3H 4R2, Canada}
\affiliation{Department of Astronomy and Physics, Saint Mary’s University, Halifax, Nova Scotia B3H 3C3, Canada}
\author{S. Costanza}
\affiliation{INFN Sezione di Pavia, I-27100 Pavia, Italy}
\author{A. Denig}
\affiliation{Institut f\"{u}r Kernphysik, Universit\"{a}t Mainz, D-55099 Mainz, Germany}
\author{M. Dieterle}
\affiliation{Departement Physik, Universit\"{a}t Basel, CH-4056 Basel, Switzerland}
\author{E. J. Downie}
\affiliation{Institut f\"{u}r Kernphysik, Universit\"{a}t Mainz, D-55099 Mainz, Germany}
\affiliation{SUPA School of Physics and Astronomy, University of Glasgow, Glasgow G12 8QQ, United Kingdom}
\affiliation{Department of Physics, The George Washington University, Washington, D.C. 20052, USA}
\author{L. V. Fil'kov}
\affiliation{Lebedev Physical Institute, 119991 Moscow, Russia}
\author{S. Garni}
\affiliation{Departement Physik, Universit\"{a}t Basel, CH-4056 Basel, Switzerland}
\author{D. I. Glazier}
\affiliation{SUPA School of Physics and Astronomy, University of Glasgow, Glasgow G12 8QQ, United Kingdom}
\affiliation{School of Physics, University of Edinburgh, Edinburgh EH9 3JZ, United Kingdom}
\author{W. Gradl}
\affiliation{Institut f\"{u}r Kernphysik, Universit\"{a}t Mainz, D-55099 Mainz, Germany}
\author{G. Gurevich}
\affiliation{Institute for Nuclear Research, 125047 Moscow, Russia}
\author{P. Hall Barrientos}
\affiliation{School of Physics, University of Edinburgh, Edinburgh EH9 3JZ, United Kingdom}
\author{D. Hamilton}
\affiliation{SUPA School of Physics and Astronomy, University of Glasgow, Glasgow G12 8QQ, United Kingdom}
\author{D. Hornidge}
\affiliation{Department of Physics, Mount Allison University, Sackville, New Brunswick E4L 1E6, Canada}
\author{D. Howdle}
\affiliation{SUPA School of Physics and Astronomy, University of Glasgow, Glasgow G12 8QQ, United Kingdom}
\author{G. M. Huber}
\affiliation{Department of Physics, University of Regina, Regina, Saskatchewan S4S 0A2, Canada}
\author{T. C. Jude}
\affiliation{School of Physics, University of Edinburgh, Edinburgh EH9 3JZ, United Kingdom}
\author{A. Kaeser}
\affiliation{Departement Physik, Universit\"{a}t Basel, CH-4056 Basel, Switzerland}
\author{V. L. Kashevarov}
\affiliation{Lebedev Physical Institute, 119991 Moscow, Russia}
\author{I. Keshelashvili}
\affiliation{Departement Physik, Universit\"{a}t Basel, CH-4056 Basel, Switzerland}
\author{R. Kondratiev}
\affiliation{Institute for Nuclear Research, 125047 Moscow, Russia}
\author{M. Korolija}
\affiliation{Rudjer Boskovic Institute, HR-10000 Zagreb, Croatia}
\author{B. Krusche}
\affiliation{Departement Physik, Universit\"{a}t Basel, CH-4056 Basel, Switzerland}
\author{A. Lazarev}
\affiliation{Joint Institute for Nuclear Research (JINR), 141980 Dubna, Russia}
\author{V. Lisin}
\affiliation{Institute for Nuclear Research, 125047 Moscow, Russia}
\author{K. Livingston}
\affiliation{SUPA School of Physics and Astronomy, University of Glasgow, Glasgow G12 8QQ, United Kingdom}
\author{I. J. D. MacGregor}
\affiliation{SUPA School of Physics and Astronomy, University of Glasgow, Glasgow G12 8QQ, United Kingdom}
\author{J. Mancell}
\affiliation{SUPA School of Physics and Astronomy, University of Glasgow, Glasgow G12 8QQ, United Kingdom}
\author{D. M. Manley}
\affiliation{Department of Physics, Kent State University, Kent, Ohio 44242, USA}
\author{W. Meyer}
\affiliation{Institut f\"{u}r Experimentalphysik, Ruhr-Universit\"{a}t, D-44780 Bochum, Germany}
\author{D. G. Middleton}
\affiliation{Institut f\"{u}r Kernphysik, Universit\"{a}t Mainz, D-55099 Mainz, Germany}
\affiliation{Department of Physics, Mount Allison University, Sackville, New Brunswick E4L 1E6, Canada}
\author{A. Mushkarenkov}
\affiliation{Department of Physics, University of Massachusetts Amherst, Amherst, Massachusetts 01003, USA}
\author{B. M. K. Nefkens}
\thanks{deceased}
\affiliation{Department of Physics and Astronomy, University of California Los Angeles, Los Angeles, California 90095-1547, USA}
\author{A. Neganov}
\affiliation{Joint Institute for Nuclear Research (JINR), 141980 Dubna, Russia}
\author{A. Nikolaev}
\affiliation{Helmholtz-Institut f\"{u}r Strahlen- und Kernphysik, Universit\"{a}t Bonn, D-53115 Bonn, Germany}
\author{M. Oberle}
\affiliation{Departement Physik, Universit\"{a}t Basel, CH-4056 Basel, Switzerland}
\author{H. Ortega Spina}
\affiliation{Institut f\"{u}r Kernphysik, Universit\"{a}t Mainz, D-55099 Mainz, Germany}
\author{M. Ostrick}
\affiliation{Institut f\"{u}r Kernphysik, Universit\"{a}t Mainz, D-55099 Mainz, Germany}
\author{P. Ott}
\affiliation{Institut f\"{u}r Kernphysik, Universit\"{a}t Mainz, D-55099 Mainz, Germany}
\author{P. B. Otte}
\affiliation{Institut f\"{u}r Kernphysik, Universit\"{a}t Mainz, D-55099 Mainz, Germany}
\author{B. Oussena}
\affiliation{Institut f\"{u}r Kernphysik, Universit\"{a}t Mainz, D-55099 Mainz, Germany}
\author{P. Pedroni}
\affiliation{INFN Sezione di Pavia, I-27100 Pavia, Italy}
\author{A. Polonski}
\affiliation{Institute for Nuclear Research, 125047 Moscow, Russia}
\author{V. Polyansky}
\affiliation{Lebedev Physical Institute, 119991 Moscow, Russia}
\author{S. Prakhov}
\affiliation{Institut f\"{u}r Kernphysik, Universit\"{a}t Mainz, D-55099 Mainz, Germany}
\affiliation{Department of Physics, The George Washington University, Washington, D.C. 20052, USA}
\affiliation{Department of Physics and Astronomy, University of California Los Angeles, Los Angeles, California 90095-1547, USA}
\author{A. Rajabi}
\affiliation{Department of Physics, University of Massachusetts Amherst, Amherst, Massachusetts 01003, USA}
\author{G. Reicherz}
\affiliation{Institut f\"{u}r Experimentalphysik, Ruhr-Universit\"{a}t, D-44780 Bochum, Germany}
\author{T. Rostomyan}
\affiliation{Departement Physik, Universit\"{a}t Basel, CH-4056 Basel, Switzerland}
\author{A. Sarty}
\affiliation{Department of Astronomy and Physics, Saint Mary’s University, Halifax, Nova Scotia B3H 3C3, Canada}
\author{S. Schrauf}
\affiliation{Institut f\"{u}r Kernphysik, Universit\"{a}t Mainz, D-55099 Mainz, Germany}
\author{S. Schumann}
\affiliation{Institut f\"{u}r Kernphysik, Universit\"{a}t Mainz, D-55099 Mainz, Germany}
\author{M. H. Sikora}
\affiliation{School of Physics, University of Edinburgh, Edinburgh EH9 3JZ, United Kingdom}
\author{A. Starostin}
\affiliation{Department of Physics and Astronomy, University of California Los Angeles, Los Angeles, California 90095-1547, USA}
\author{O. Steffen}
\affiliation{Institut f\"{u}r Kernphysik, Universit\"{a}t Mainz, D-55099 Mainz, Germany}
\author{I. I. Strakovsky}
\affiliation{Department of Physics, The George Washington University, Washington, D.C. 20052, USA}
\author{T. Strub}
\affiliation{Departement Physik, Universit\"{a}t Basel, CH-4056 Basel, Switzerland}
\author{I. Supek}
\affiliation{Rudjer Boskovic Institute, HR-10000 Zagreb, Croatia}
\author{M. Thiel}
\affiliation{II. Physikalisches Institut, Universit\"{a}t Giessen, D-35392 Giessen, Germany}
\author{L. Tiator}
\affiliation{Institut f\"{u}r Kernphysik, Universit\"{a}t Mainz, D-55099 Mainz, Germany}
\author{A. Thomas}
\affiliation{Institut f\"{u}r Kernphysik, Universit\"{a}t Mainz, D-55099 Mainz, Germany}
\author{M. Unverzagt}
\affiliation{Institut f\"{u}r Kernphysik, Universit\"{a}t Mainz, D-55099 Mainz, Germany}
\affiliation{Helmholtz-Institut f\"{u}r Strahlen- und Kernphysik, Universit\"{a}t Bonn, D-53115 Bonn, Germany}
\author{Y. Usov}
\affiliation{Joint Institute for Nuclear Research (JINR), 141980 Dubna, Russia}
\author{D. P. Watts}
\affiliation{School of Physics, University of Edinburgh, Edinburgh EH9 3JZ, United Kingdom}
\author{L. Witthauer}
\affiliation{Departement Physik, Universit\"{a}t Basel, CH-4056 Basel, Switzerland}
\author{D. Werthm\"{u}ller}
\affiliation{Departement Physik, Universit\"{a}t Basel, CH-4056 Basel, Switzerland}
\author{M. Wolfes}
\affiliation{Institut f\"{u}r Kernphysik, Universit\"{a}t Mainz, D-55099 Mainz, Germany}

\collaboration{A2 Collaboration at MAMI}\noaffiliation

\date{\today}

\begin{abstract}
The spin polarizabilities of the nucleon describe how the spin of the nucleon responds to an incident polarized photon.  The most model-independent way to extract the nucleon spin polarizabilities is through polarized Compton scattering. 
Double-polarized Compton scattering asymmetries on the proton were measured in the $\Delta(1232)$ region using circularly polarized incident photons and a transversely polarized proton target at the Mainz Microtron. 
Fits to asymmetry data were performed using a dispersion model calculation and a baryon chiral perturbation theory calculation, and a separation of all four proton spin polarizabilities in the multipole basis was achieved.
The analysis based on a dispersion model calculation yields $\gamma_{E1E1} = -3.5 \pm 1.2$, $\gamma_{M1M1}= 3.16 \pm 0.85$, $\gamma_{E1M2} = -0.7 \pm 1.2$, and $\gamma_{M1E2} = 1.99 \pm 0.29$, in units of $10^{-4}$ fm$^4$.

\end{abstract}


\maketitle


Electromagnetic polarizabilities are fundamental properties of composite systems such as molecules, atoms, nuclei, and hadrons \cite{Hol92}.
Whereas magnetic moments provide information about the ground-state properties of a system, polarizabilities provide information about the excited states of the system.
For atomic systems, polarizabilities are of the order of the atomic volume.
For hadrons, polarizabilities are much smaller than the volume, typically of order $10^{-4}$ fm$^3$, because of the greater strength of the QCD force as compared to the electromagnetic force.
Extracted polarizabilities can provide a guide by favoring, or disfavoring, models of hadron structure and QCD.

Hadron polarizabilities are best extracted with Compton scattering experiments, where the polarizabilities cause a deviation of the cross section from the prediction of Compton scattering off a structureless Dirac particle.
In the energy expansion of the nuclear Compton scattering amplitude, the $\mathcal{O}(\omega^{2})$ term depends on the electric and magnetic, or scalar, polarizabilities of the nucleon, $\alpha$ and $\beta$, respectively, and the $\mathcal{O}(\omega^{3})$ term depends on the spin polarizabilities of the nucleon, where $\omega$ is the incident photon energy. The $\mathcal{O}(\omega^{3})$ term in the amplitude is \cite{Bab98} an effective spin-dependent interaction,
\begin{flalign}\label{Eq:Comp3}
H_{\mathrm{eff}}^{(3)} = -4\pi\!\bigg[&\frac{1}{2}\gamma_{E1E1}\vec{\sigma}\cdot(\vec{E}\times\dot{\vec{E}})+\frac{1}{2}\gamma_{M1M1}\vec{\sigma}\cdot(\vec{H}\times\dot{\vec{H}}) \nonumber \\
&-\gamma_{M1E2}E_{ij}\sigma_{i}H_{j}+\gamma_{E1M2}H_{ij}\sigma_{i}E_{j}\bigg]\text{,}
\end{flalign}
which describes the coupling of the proton spin, $\vec{\sigma}$, with an applied electric $\vec{E}$ or magnetic $\vec{H}$ field and their time derivatives, $\dot{\vec{E}}=\partial_{t}\vec{E}$, and space derivatives, $E_{ij}=\frac{1}{2}\left(\nabla_{i}E_{j}+\nabla_{j}E_{i}\right)$, in the multipole basis, where $\gamma_{X\lambda Y\lambda^{\prime}}$ is the spin polarizability for incident and final photon multipolarities $X\lambda$ and $Y\lambda^{\prime}$.
Because the spin polarizability effect varies as $\omega^{3}$, the sensitivity to the spin polarizabilities, relative to that of $\alpha$ and $\beta$, is greatest in Compton scattering reactions in the $\Delta(1232)$ region, but below the threshold for double-pion photoproduction where additional terms complicate matters.

Several experiments have provided constraints on linear combinations of the proton spin polarizabilities.
The most important of these are (i) the forward spin polarizability $\gamma_{0}$, which comes from a set of two experiments of the GDH Collaboration \cite{Ahr01,Dut03}, 
$\gamma_{0}=-\gamma_{E1E1}-\gamma_{E1M2}-\gamma_{M1E2}-\gamma_{M1M1} =(-1.01 \pm 0.08 \pm 0.10)\times10^{-4}\,\text{fm}^{4}$, and (ii) the backward spin polarizability $\gamma_{\pi}$, which was determined from an analysis of backward angle Compton scattering \cite{Cam02}, 
$\gamma_{\pi}=-\gamma_{E1E1}-\gamma_{E1M2}+\gamma_{M1E2}+\gamma_{M1M1} =(8.0 \pm 1.8)\times10^{-4}\,\text{fm}^{4}$.
The convention followed here is to subtract the structureless pion-pole contribution from the spin polarizability; the pole term is present in $\gamma_{\pi}$ and the multipole basis spin polarizabilities \cite{Bab98}, but is not present in $\gamma_{0}$. 
Table \ref{Tab:SPs} presents the results from several theoretical calculations for the spin polarizabilities, showing the wide range of theoretical predictions. 

\begin{table*}[!ht]
\begin{tabular}{|c|c|c|c|c|c|c|c|c|c|c|}
\hline
& $O(\epsilon^3)$ & $O(p^{4})_{a}$ & $O(p^{4})_{b}$ & $K$-matrix & HDPV & DPV & $L_{\chi}$ & HB$\chi$PT & B$\chi$PT & Experiment \\ \hline

$\gamma_{E1E1}$ & $-1.9$ & $-5.4$ & 1.3 & $-4.8$ & $-4.3$ & $-3.8$ & $-3.7$ & $-1.1 \pm 1.8$ (theory) & $-3.3$ & $-3.5 \pm 1.2$ \\ \hline

$\gamma_{M1M1}$ & 0.4 & 1.4 & 3.3 & 3.5 & 2.9 & 2.9 & 2.5 & $2.2 \pm 0.5$ (stat) $\pm$ 0.7 (theory) & 3.0 & $3.16 \pm 0.85$ \\ \hline

$\gamma_{E1M2}$ & 0.7 & 1.0 & 0.2 & $-1.8$ & $-0.02$ & 0.5 & 1.2 & $-0.4 \pm 0.4$ (theory) & 0.2 & $-0.7 \pm 1.2$\\ \hline

$\gamma_{M1E2}$ & 1.9 & 1.0 & 1.8 & 1.1 & 2.2 & 1.6 & 1.2 & $1.9 \pm 0.4$ (theory) & 1.1 & $1.99 \pm 0.29$ \\ \hline

$\gamma_{0}$ & $-1.1$ & 1.9 & $-3.9$ & 2.0 & $-0.8$ & $-1.1$ & $-1.2$ & $-2.6$ & $-1.0$ & $-1.01 \pm 0.08 \pm 0.10$ \cite{Ahr01,Dut03} \\ \hline

$\gamma_{\pi}$ & 3.5 & 6.8 & 6.1 & 11.2 & 9.4 & 7.8 & 6.1 & 5.6 & 7.2 & $8.0 \pm 1.8$ \cite{Cam02} \\ \hline

\end{tabular}
\caption[Predictions and experiment for proton spin polarizabilities]{Spin polarizabilities in units of $10^{-4}\,\text{fm}^{4}$.
$O(\epsilon^3)$ is a small scale expansion calculation \cite{Hem98}. $O(p^4)_{a,b}$ are chiral perturbation theory calculations \cite{Kum00,Gel01}.
The $K$-matrix calculation is from Ref.\ \cite{Kon01}.
HDPV and DPV are fixed-$t$ \cite{Hol00,Pas07} and fixed-angle \cite{Dre03} dispersion relation calculations, respectively, where the acronyms represent the authors of the respective papers.
$L_{\chi}$ is a chiral Lagrangian calculation \cite{Gas11}.
HB$\chi$PT and B$\chi$PT are heavy baryon and covariant, respectively, chiral perturbation theory calculations with $\Delta(1232)$ degrees of freedom \cite{McG13,Len14}.
Experimental results for $\gamma_{E1E1}$, $\gamma_{E1M2}$, $\gamma_{M1E2}$, and $\gamma_{M1M1}$ are from this work, using a combined analysis of $\Sigma_{2x}$ and $\Sigma_{3}$ asymmetries using a dispersion model calculation \cite{Dre03}.}
\label{Tab:SPs}
\end{table*}

Compton scattering asymmetries in the $\Delta(1232)$ region have sensitivity to the spin polarizabilities \cite{Pas07}, with the relationship described in Eqs.\ (3.23), (3.26), and (3.15) of Ref.\ \cite{Bab98}.
This Letter presents the first measurements of double-polarized Compton scattering asymmetries on the nucleon at energies below the double-pion photoproduction threshold, and the first analysis of Compton scattering asymmetries for the determination of all four spin polarizabilities. 
The double-polarization asymmetry with circularly polarized incident photons on a transversely polarized proton target $\Sigma_{2x}$ was measured using the Crystal Ball (CB) detector \cite{Sta01} at the Mainz Microtron (MAMI) \cite{Kai08}.
A GaAsP (III-V semiconductor) source was used to produce a longitudinally polarized electron beam, with the polarization measured via a Mott polarimeter \cite{Tio11}.
The average beam polarization was $81.9 \pm 0.1\%$.
To remove systematic effects, the helicity of the beam was automatically flipped at a frequency of about 1 Hz.
The 450 MeV electron beam produced by the MAMI accelerator was passed through a 10 $\mu$m Cu radiator, producing a circularly polarized bremsstrahlung photon beam.
The energy of the radiated photon was determined via the detection of the scattered electron in the Glasgow photon tagger \cite{McG08}, and only photons in the range  $E_{\gamma}=273$--$303$ MeV were used in this analysis.
After collimation by a 2.5 mm diameter lead collimator, the photon beam impinged on a frozen spin butanol target \cite{Tho11}.
The target was polarized by dynamic nuclear polarization \cite{Cra97}, typically up to initial values of 90\% with relaxation times on the order of 1000 h.
Proton polarizations were measured using a NMR coil at the beginning and end of a polarization period, with an average of $81.6 \pm 1.7\%$.
To further remove systematic effects, the direction of proton polarization was reversed several times, typically once per week of experiment running time.  

To remove backgrounds from interactions of the photon beam with the material of the cryostat and nonhydrogen nucleons in the target and He bath, separate data were taken using a carbon foam target with density 0.55 g/cm$^3$.
The density of the carbon foam was such that a cylinder of identical geometric size to the butanol target provided a close approximation to the number of nonhydrogen nucleons in the butanol target, allowing for a simple 1:1 subtraction accounting only for differences in luminosity. 

Final-state particles were detected in the CB \cite{Sta01} and TAPS \cite{Nov91} detectors, both of which are outfitted with charged particle identification systems \cite{Tar08}.
Together these detectors cover 97\% of $4\pi$ sr.
Events were selected where a single neutral and a single charged cluster of detector element hits, both with energies above 15 MeV, were observed in coincidence with an event in the photon tagger.
A prompt timing selection was applied followed by an accidental coincidence subtraction.
An additional level of background suppression was achieved by requiring a projected angle of less than 10$^{\circ}$ between the measured direction of the charged particle and the direction of the proton recoil predicted by the Compton scattering kinematics.
The 10$^{\circ}$ opening angle requirement was determined through simulation and checked with $\pi^{0}$ photoproduction data.

\begin{figure}
  \includegraphics[width=3.4in]{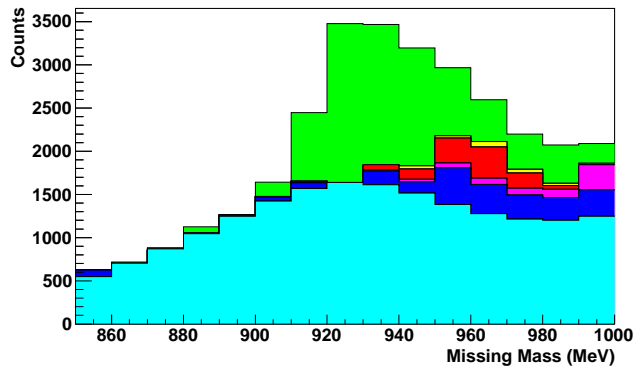}
  \caption{(color online). Missing-mass spectra for $E_{\gamma} = 273$--$303$ MeV, and $\theta_{c} = 100^{\circ}$--$120^{\circ}$. In addition to the actual Compton scattering distribution, each background, shown in a different color, is stacked on top of one another to show its contribution to the total initial distribution. From bottom to top; light blue is for tagger accidentals; blue is for carbon or cryostat background; magenta, red, and yellow were constructed from data to mimic where a $\pi^{0}$ decay photon was lost in the upstream CB hole, the region between CB and TAPS, and the downstream TAPS hole, respectively, each of which had their own accidental and carbon subtraction already applied; and green shows the final subtracted result.}
  \label{Fig:MMaWith}
\end{figure}

The ratio of $\pi^{0}$ photoproduction to Compton scattering cross sections in the $\Delta(1232)$ region is approximately 100:1.
Even with the exclusivity selection, accidental subtraction, and opening angle requirement, $\pi^{0}$ backgrounds remained in the data, as shown in the missing-mass spectrum of Fig.\ \ref{Fig:MMaWith}.
Typically, these backgrounds were from $\pi^{0}$ events in which a low-energy decay photon escaped detection by passing up or down the beam line, or through the gap between the CB and TAPS.
To isolate this background, selections were used to make the regions of reduced acceptance well defined.
These regions were (i) the forward hole in the TAPS detector, $0^{\circ}$--$6^{\circ}$, (ii) the region between TAPS and the CB, $18^{\circ}$--$25^{\circ}$, and (iii) the backward hole in the CB, $150^{\circ}$--$180^{\circ}$.
This aggregate solid angle is referenced as $\Omega_{\text{cut}}$.
To estimate the $\pi^{0}$ background that resulted from decay photons entering $\Omega_{\text{cut}}$, $\pi^{0}$ events were identified where both decay photons were detected, and where one of the decay photons fell into a solid angle bin adjacent to $\Omega_{\text{cut}}$.
For $\Omega_{\text{cut}}=0^{\circ}$--$6^{\circ}$, the adjacent solid angle bin is $\Omega_{\text{adj}}=6^{\circ}$--$8.5^{\circ}$; for $\Omega_{\text{cut}}=18^{\circ}$--$25^{\circ}$, $\Omega_{\text{adj}}=13.2^{\circ}$--$27.9^{\circ}$; and for $\Omega_{\text{cut}}=150^{\circ}$--$180^{\circ}$, $\Omega_{\text{adj}}=143^{\circ}$--$150^{\circ}$.
Missing-mass spectra were calculated using
\begin{equation}\label{Eq:MM}
M_{\text{miss}}=\sqrt{(E_{\gamma}+m_{p}-E_{c})^{2}-(\vec{p}_{\gamma}-\vec{p}_{c})^{2}}\text{,}
\end{equation}
where $E_{c}$ and $\vec{p}_{c}$ are the energy and momentum of the Compton photon.
For the $\pi^{0}$ background events the photon detected in $\Omega_{\text{adj}}$ was ignored, and the second photon was treated as the Compton photon.
 
\begin{figure}
  \includegraphics[width=3.4in]{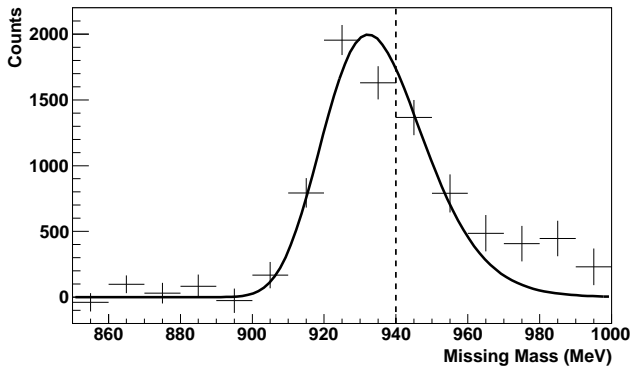}
  \caption{Missing-mass spectrum after removing the background contributions shown in Fig.\ \ref{Fig:MMaWith} for $E_{\gamma} = 273$--$303$ MeV, and $\theta_{c} = 100^{\circ}$--$120^{\circ}$. The solid line is the Compton scattering line shape determined from simulation. The dashed line indicates the upper integration limit used in the analysis.}
  \label{Fig:MMaSubt}
\end{figure}

After removing the various background contributions shown in Fig.\ \ref{Fig:MMaWith} the final missing-mass distribution is shown in Fig.\ \ref{Fig:MMaSubt}.
The subtraction of backgrounds is done separately for each helicity state, as the $\pi^{0}$ backgrounds themselves result in nonzero asymmetries.
Monte Carlo simulation of the Compton scattering line shape shows good agreement between data and calculation for the Compton peak, except around 980 MeV.
The counts there are from an unsubtracted background due to the gap between the CB and TAPS which simulation shows would appear on the high $M_{\text{miss}}$ side of the peak.
While relatively weak at $\theta_{c} = 100^{\circ}$--$120^{\circ}$, this background becomes stronger at more backward angles.
The effect of the background on the asymmetry was studied by integrating the spectrum in Fig.\ \ref{Fig:MMaSubt} to various upper limits.
Integrating to final values up to 940 MeV resulted in asymmetries consistent within uncertainties, while integrating to increasingly higher values resulted in asymmetries that varied outside of uncertainties from lower limit integrations.
For this reason a relatively conservative integration limit of 940 MeV was used in this analysis.
Further details of this analysis can be found in Ref.\ \cite{Mar13}

For a given incoming photon energy $E_{\gamma}$, Compton scattering polar angle $\theta_{c}$, and azimuthal angle $\phi_{c}$ relative to the target polarization direction, the asymmetry $\Sigma_{2x}$ is defined by
\begin{flalign}\label{Eq:Sig2xP}
  \Sigma_{2x}(E_{\gamma},\theta_{c})=&\left[P_{T}P_{\gamma}(E_{\gamma})\mathrm{cos}(\phi_{c})\right]^{-1}\nonumber \\
&\left[\frac{N^{R}(E_{\gamma},\theta_{c},\phi_{c})-N^{L}(E_{\gamma},\theta_{c},\phi_{c})}{N^{R}(E_{\gamma},\theta_{c},\phi_{c})+N^{L}(E_{\gamma},\theta_{c},\phi_{c})}\right]\text{,}
\end{flalign}
where $P_{T}$ is the target polarization, $P_{\gamma}$ is the beam polarization, and $N^{R}$ ($N^{L}$) are the counts in the specified bin with a right (left) helicity beam.

\begin{figure}
  \includegraphics[width=3.4in]{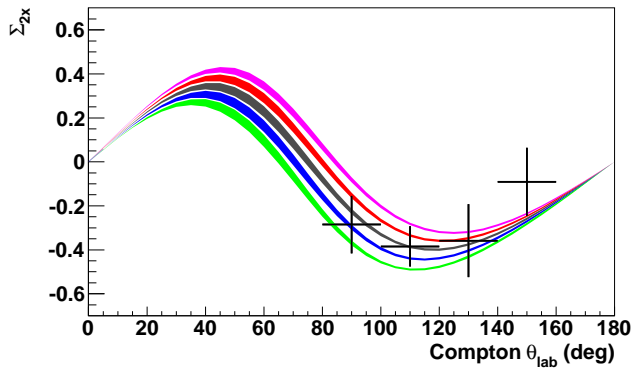}
  \caption{(color online). $\Sigma_{2x}$ for $E_{\gamma} = 273$--$303$ MeV. The curves are from a dispersion theory calculation \cite{Dre03} with $\alpha$, $\beta$, $\gamma_{0}$, and $\gamma_{\pi}$ held fixed at their experimental values, and $\gamma_{M1M1}$ fixed at $2.9$. From bottom to top, the green, blue, brown, red, and magenta bands are for $\gamma_{E1E1}$ equal to $-6.3$, $-5.3$, $-4.3$, $-3.3$, and $-2.3$, respectively. The width of each band represents the propagated errors from $\alpha$, $\beta$, $\gamma_{0}$, and $\gamma_{\pi}$ combined in quadrature.}
\label{Fig:AsyE1}
\end{figure}

The measured asymmetries are plotted in Fig.\ \ref{Fig:AsyE1}.
In addition to statistical uncertainties, the systematic uncertainties from both beam and target polarizations are incorporated in the error bars shown, though at worst they are only 9\% of the total uncertainty.
Systematic uncertainty from the carbon background subtraction was estimated by varying the carbon background ratio by $\pm 20\%$ from the expected value.
The effect on the asymmetries was negligible, at worst about 10\% of the total uncertainty, and this systematic uncertainty is not included in the error bars shown in Fig.\ \ref{Fig:AsyE1}.
The curves are from a dispersion theory calculation \cite{Dre03} for values of $\gamma_{E1E1}$ ranging from $-6.3$ to $-2.3$, but with $\gamma_{M1M1}$ fixed at the HDPV value from Table \ref{Tab:SPs} of $2.9$ \cite{Hol00,Pas07}.
The width of each band represents the propagated errors using $\alpha = 12.16 \pm 0.58$ and $\beta = 1.66 \pm 0.69$, as well as $\gamma_{0}$ and $\gamma_{\pi}$ from Table \ref{Tab:SPs}, combined in quadrature.
The curves graphically demonstrate the sensitivity of the asymmetries to $\gamma_{E1E1}$, showing a preferred solution of $\gamma_{E1E1} \approx -4.3 \pm 1.5$.
A similar analysis holding $\gamma_{E1E1} = -4.3$ fixed and allowing $\gamma_{M1M1}$ to vary shows that the asymmetries are insensitive to $\gamma_{M1M1}$.

The double-asymmetry data from this measurement, and published results \cite{Leg01} for the single-polarization asymmetry with linearly polarized photons $\Sigma_{3}$, were fitted with a dispersion model calculation \cite{Dre03} and a baryon chiral perturbation theory (B$\chi$PT) calculation \cite{Len10}.
Only asymmetry points obtained by the LEGS Collaboration below double-pion photoproduction threshold were used in this analysis.
The B$\chi$PT calculation includes pion, nucleon, and $\Delta(1232)$ degrees of freedom at next-to-next-to-leading order \cite{Len10}.
$\alpha$, $\beta$, $\gamma_{E1E1}$, $\gamma_{M1M1}$, $\gamma_{0}$, and $\gamma_{\pi}$ were fitted to the asymmetry data sets, and to the known constraints on $\alpha+\beta$, $\alpha-\beta$, $\gamma_{0}$, and $\gamma_{\pi}$.
The constraint on $\alpha + \beta$ is given by the Baldin sum rule, $\alpha + \beta = 13.8 \pm 0.4$ \cite{Olm01}, and the constraint of $\alpha - \beta = 7.6 \pm 0.9$ is taken from the analysis of Grie{\ss}hammer \emph{et al.} \cite{Gri12}.

\begin{table}[!ht]
  \begin{tabular}{|c|c|c|c|}
    \hline
    Data fit & Model & $\gamma_{E1E1}$ & $\gamma_{M1M1}$ \\ \hline
    $\Sigma_{2x}$ & Disp & $-4.6 \pm 1.6$ & $-7 \pm 11$ \\ \hline
    $\Sigma_{3}$  & Disp & $-1.4 \pm 1.7$ & $3.20 \pm 0.85$ \\ \hline
    $\Sigma_{2x}$ and $\Sigma_{3}$ & Disp  & $-3.5 \pm 1.2$ & $3.16 \pm 0.85$ \\ \hline
    $\Sigma_{2x}$ and $\Sigma_{3}$ & B$\chi$PT & $-2.6 \pm 0.8$ & $2.7 \pm 0.5$ \\ \hline
  \end{tabular}
  \caption{Results from fitting $\Sigma_{2x}$ (this work) and $\Sigma_{3}$ \cite{Leg01} data using either a dispersion model calculation (Disp) \cite{Dre03} or a B$\chi$PT calculation \cite{Len10}.}
  \label{Tab:Fit}
\end{table}

Table \ref{Tab:Fit} shows results from data fitting.
The first column gives the data set used for fitting, the second column gives the model used, and the third and fourth columns show the results for $\gamma_{E1E1}$ and $\gamma_{M1M1}$.   
The first row shows results from fitting only the $\Sigma_{2x}$ data from this work.
The result for $\gamma_{E1E1}$ is in good agreement with the expectation from the graphical analysis shown in Fig.\ \ref{Fig:AsyE1}, and the $\Sigma_{2x}$ data alone have little sensitivity to $\gamma_{M1M1}$.
The second row shows results from fitting only $\Sigma_{3}$ \cite{Leg01}.
Within uncertainties, the results for $\gamma_{E1E1}$ from fitting $\Sigma_{2x}$ and $\Sigma_{3}$ data separately are in approximate agreement.
The third row shows the results from the combined fit of $\Sigma_{2x}$ and $\Sigma_{3}$ using the dispersion model \cite{Dre03}, and the fourth row shows the combined fit of $\Sigma_{2x}$ and $\Sigma_{3}$ using the B$\chi$PT calculation \cite{Len10}.
Within uncertainties, the results for $\gamma_{E1E1}$ and $\gamma_{M1M1}$ from the two models are also in agreement.
This indicates that the model dependence of the polarizability fitting is comparable to, or smaller than, the statistical errors from data fitting.

Results for all four spin polarizabilities obtained from the combined fit of $\Sigma_{2x}$ and $\Sigma_{3}$ using the dispersion model calculation are presented in the last column in Table \ref{Tab:SPs}, along with previous results for $\gamma_{0}$ and $\gamma_{\pi}$.
$\gamma_{E1M2}$ and $\gamma_{M1E2}$ were extracted using the linear relationships of $\gamma_{0}$ and $\gamma_{\pi}$.
The table shows generally good agreement between the extracted spin polarizabilities and the predictions of the dispersion theory calculations \cite{Hol00,Pas07,Dre03}, the $K$-matrix theory calculation \cite{Kon01}, the chiral Lagrangian calculation \cite{Gas11}, and the chiral perturbation theory calculations \cite{McG13,Len14}.
The size of the experimental uncertainties is too large to discriminate between these various models.

In summary, data are presented for the double-polarized Compton scattering asymmetry with a transversely polarized proton target in the $\Delta(1232)$ region.
The data have good sensitivity to the $\gamma_{E1E1}$ spin polarizability.
The spin polarizabilities obtained using the dispersion theory analysis \cite{Dre03} of the asymmetry data, and those obtained using the B$\chi$PT analysis \cite{Len10} of the data, agree within uncertainties. 
The spin polarizabilities are in good agreement with the dispersion theory, $K$-matrix theory, the chiral Lagrangian calculation, and the chiral perturbation theory calculations. 


We wish to acknowledge the outstanding support of the accelerator group and operators of MAMI.
We also wish to acknowledge and thank D. Drechsel, H. Grie{\ss}hammer, B. Holstein, J. McGovern, V. Pascalutsa, B. Pasquini, D. Phillips, and M. Vanderhaeghen for their comments and suggestions.
This material is based on work supported by the U.S. Department of Energy Office of Science, Office of Nuclear Physics, under Awards No. DE-FG02-88-ER40415, No. DE-FG02-99-ER41110, and No. DE-FG02-01-ER41194, the National Science Foundation under Grants No. PHY-1039130 and No. IIA-1358175, the Deutsche Forschungsgemeinschaft (SFB443, SFB/TR16, and SFB1044), DFG-RFBR (Grants No. 09-02-91330 and No. 05-02-04014), the European Community-Research Infrastructure Activity (FP6), Schweizerischer Nationalfonds, the UK Science and Technology Facilities Council (STFC 57071/1, 50727/1), INFN (Italy), and NSERC (Canada).


\bibliography{pmartel_spinpol}

\end{document}